\documentclass[prx,aps,english,twocolumn,notitlepage,superscriptaddress]{revtex4-2}
\usepackage[T1]{fontenc}  
\usepackage[latin1]{inputenc}
\usepackage{multirow,amsmath}
\usepackage{babel} 
\usepackage{txfonts}
\usepackage{epsfig,epsf,psfrag} 
\usepackage{graphicx} 
\usepackage{subfigure}
\usepackage{pslatex}
\usepackage{float}
\usepackage{upgreek}
\usepackage{subfigure}
\usepackage{epic,eepic} 
\usepackage{color,pstcol}
\usepackage{pstricks} 
\usepackage{fancyhdr} 
\usepackage{tabularx}
\usepackage{physics}
\usepackage{url}
\usepackage{listings}
\usepackage{color}
\begin{document}
\lstset{language=Mathematica}
\title{Testing quantum speedups in exciton transport through a photosynthetic complex using quantum stochastic walks}
\author{Naini Dudhe}\email{nainidudhe@gmail.com}
\affiliation{School of Physical Sciences, National Institute of Science Education \& Research, HBNI, Jatni-752050, India}
\author{Pratyush Kumar Sahoo}\email{pratyushkumar.iiser.gmail.com}
\affiliation{Department of Physical Sciences, Indian Institute of Science Education \& Research, Kolkata, India}
\author{Colin Benjamin}\email{colin.nano@gmail.com}
\affiliation{School of Physical Sciences, National Institute of Science Education \& Research, HBNI, Jatni-752050, India} 
\begin{abstract}
Photosynthesis is a highly efficient process, nearly $100$ percent of the red photons falling on the surface of leaves reach the reaction center and get transformed into energy. Most theoretical studies  on photosynthetic complexes focus mainly on the  Fenna-Matthews-Olson complex obtained from green-sulfur bacteria.  Quantum coherence was speculated to play a significant role in this very efficient transport process. However, recent reports indicate quantum coherence via exciton transport may not be as relevant as coherence originating via vibronic processes to Photosynthesis. Regardless of the origin, there has been a debate on whether quantum coherence results in any speedup of the exciton transport process. To address this we  model exciton transport in FMO using a quantum stochastic walk (QSW) with only incoherence, pure dephasing and with both dephasing and incoherence. We find that the QSW model with pure dephasing leads to a substantial speedup in exciton transport as compared to a QSW model which includes both dephasing and incoherence and one which includes only incoherence, both of which experience slowdowns.
\end{abstract}
\maketitle
\section{\label{Introduction}Introduction}
The first step  of photosynthesis takes place via the antennae molecules called the chromophores. These antennae molecules lose electrons when light falls on them and an electron-hole pair or exciton is formed~\cite{Jim}. This exciton in turn has to reach the reaction center where the charge separation occurs and energy is stored. Usually these reaction centers are far in terms of molecular distance from the excited antennae molecule. But, this process of transferring the captured photon to the reaction centre is seen to exhibit an efficiency close to $100$ percent. In $2007$, it was reported~\cite{Engel_2010} using ``two dimensional Fourier transform electronic spectroscopy'' (2D-FTES) that quantum coherence could be playing an important role in the exciton transport in Fenna-Matthews-Olson (FMO) complex found in green-sulphur bacteria. Later this was theoretically analyzed~\cite{41}. However, more recently the origin of this quantum coherence in FMO complexes is speculated to arise via vibrational coupling of excitons to the bath~\cite{cao, brumer}.

Regardless of the origin whether vibronic or excitonic, the efficiency of the photosynthetic process has been a mystery. Speculation has been rife that the exciton may not be following a classical random walk to get to the reaction center before its conversion to energy~\cite{Jim}. Rather the antennae molecules could be operating via a search strategy called the quantum walk~\cite{41}. A quantum walker takes all possible paths, (like a superposed atom in the two-slit experiment) as opposed to classical random walker who must choose a single route. This gives quantum walk an advantage in the sense that it spreads with rate proportional to the time taken as compared to classical random walk which spreads as square root of time. The exciton transfer seen at life sustaining temperatures of $300~K$ is thus aided via a quantum walk process which apparently takes place in presence of dephasing and/or incoherence.\\
In Ref.~\cite{41} quantum walks were first used to study exciton transfer dynamics in FMO complex interacting with a thermal bath. Later, Hoyer, et. al., in Ref.~\cite{Hoyer_2010} used a master equation approach with only incoherence to model exciton transfer in FMO. In Ref.~\cite{QSW_def} a more general quantum walk called quantum stochastic walk (QSW) was introduced. QSW is versatile and includes classical effects, either only dephasing or only incoherence or both incoherence and dephasing. QSW interpolates between classical random walk (CRW) and continuous-time quantum walk (CTQW) through a single parameter $\omega$. $\omega$ is a measure of the amount of dephasing and/or incoherence built into the QSW. Later in Ref.~\cite{QSWalk}, QSW was used to model FMO and results were compared with Ref.~\cite{Hoyer_2010}. Although natural light harvesting complexes such as the chlorosome antenna complex in green sulphur bacteria are known to exhibit non-Markovianity~\cite{liu2015quantifying,zheng2021fully,marquez2016probing}, we are justified in using the Markovian approach. This is so because we compare our results with those of Ref.~\cite{Hoyer_2010}, which relies on the Born-Markov approximation. Although irrelevant to our work, it is important to note that QSW can be extended to the non-Markovian regime as well, see chapter 5 of Ref.~\cite{schijven2014quantum} and Ref.~\cite{schijven2013energy}. Our main aim in this paper is to test the prognosis that quantum effects do not lead to any speedup of the exciton transport from antenna to reaction center as was advanced in Ref.~\cite{Hoyer_2010} as well as Ref.~\cite{nike}. In the latter too, it was observed that both the incoherent approach and the approach fully taking into consideration quantum coherence, saw no speed up in exciton transport from antenna to reaction center. Also, if there is indeed no speedup in the only incoherent scattering model, does there exist a scheme which might lead to a quantum speedup? Quantum speedup of excitonic transport is measured via the localization time ($t_{loc}$). Localization time~\cite{Hoyer_2010} is defined as time at which the onset of sub-diffusive transport occurs in the exciton transport process. The greater is the localization time, the more is the duration for which super-diffusive transport prevails. Thus, for significant quantum speed up, the localization time must be large. In Ref.~\cite{Hoyer_2010} localization time at both $77~K$ and $300~K$ is around $70~fs$. In our work, we find for the QSW model with only incoherence, localization time at $300~K$ is more than that at $77~K$ in accordance with the expectation of a slowdown. The main take home message of this work is that at life sustaining temperatures of $300~K$, the QSW model with only incoherence leads to a slowdown in exciton transport similar to the case of a QSW model with both dephasing and incoherence, in contrast to the master equation approach of Hoyer, et. al. which shows no change in $t_{loc}$. In Ref.~\cite{Hoyer_2010} there is no quantum speedup seen at $300~K$. Moreover, the results of the QSW model with both dephasing and incoherence are in line with another recent study (includes experimental as well as theoretical analysis using non-local quantum master equation approach) which debunks any quantum coherent effects in photosynthetic transport~\cite{duan}. Thus, our work mainly focuses on two points. The first would be using our model with only incoherent scattering scheme and comparing our results with Ref.~\cite{Hoyer_2010} which also uses the same scheme. We observe a difference from Ref.~\cite{Hoyer_2010} and see a significant slowdown for both dephasing with incoherence and only incoherent transport mechanisms. Secondly, we find a scheme which shows a speedup, this is the pure dephasing transport scheme.

The outline of the paper is as follows: in the next section we give details of the quantum stochastic walk used to model exciton transport in FMO and introduce the three models, one incorporating only incoherence, the other pure dephasing and finally, one including both incoherence and dephasing. Subsequent to this we give details of the FMO complex, its Hamiltonian and how the QSW schemes are used to model exciton transport in FMO. In section III we plot the results of our simulations for total site coherence, site population, mean square displacement and localization time. In section IV we discuss our results and plots via two tables, the first for localization time and the second for other quantities. We end the paper and section IV with conclusion which includes a perspective on future endeavors in this area. At the end, we close with an appendix which provides the codes and explains the methdology behind our calculations.
\section{Quantum Stochastic Walk}
Two variants of quantum walks are known- discrete-time quantum walk (DTQW)~\cite{coineddef} and continuous time quantum walk (CTQW)~\cite{Farhi}. A more general form of continuous time quantum walk called the quantum stochastic walk (QSW) was first introduced in Ref.~\cite{QSW_def}. It has the advantage of interpolating continuously from a classical random walk to a continuous time quantum walk and can address quantum walk processes which are coupled to an environment. QSW was derived from Kossakowski-Lindblad master equation~\cite{KOSSAKOWSKI,Lindblad}, which is used to describe quantum stochastic process and model open quantum system dynamics. QSW is based on density matrix, dynamics of which is given by~\cite{QSW_def}-
\begin{equation}
 \begin{array}{ccl}
    \frac{d\rho(t)}{dt}&=&-\left(1-\omega \right)i\left[H,{\rho(t)}\right]\\
    & & +\omega\sum_{k=1}^{K} \left(\hat{L}_{k}\rho(t)\hat{L}_{k}^{\dagger} -\frac{1}{2}\left(\hat{L}_{k}^{\dagger}\hat{L}_{k}\rho(t)+\rho(t)\hat{L}_{k}^{\dagger}\hat{L}_{k}\right)\right),
    \label{QWS_def}
\end{array}
\end{equation}
where $\rho(t)$ is the density matrix representation of walker at time $t$. For QSW on  any graph G, $\rho(t)$ is
a $N \times N$ matrix with vertex states- $\{\ket{1}, . . . , \ket{N}\}$ as the basis, and elements $\rho_{ij}(t) = \bra{i} \rho(t) \ket{j}$. $0\leq \omega\leq 1$ is a parameter which interpolates between CRW and CTQW. $H$ is the Hamiltonian operator responsible for coherent dynamics. The index $k$ has a unique value for each pair of $i,j$. The summation over $k$ is from 1 to $K$ where $K$ depends on the choice of the Lindblad operator. The set of Lindblad operators essentially represents scattering between each pair of vertices. For the dephasing with incoherence scheme, $K=N^2$ since there are $N^2$ combinations in order to incorporate any permutation of two vertices out of $N$. For this case, both $i$ and $j$ run over all vertices, hence, they can have $N^2$ combinations. On the other hand, for the pure dephasing scattering scheme, the Lindblad operator involves only the diagonal elements of the Hamiltonian, thus $K=N$ for this case. Similarly, for the case of only incoherent scattering, $K=N(N-1)$ since only the off-diagonal elements are taken into account (see Ref.~\cite{QSWalk}). The first term on the right hand side of Eq.~(\ref{QWS_def}) is responsible for coherent dynamics and the second term on right gives rise to incoherent dynamics via only dephasing, both dephasing and incoherent scattering or only incoherent scattering. $\hat{L}_{k}$ are Lindblad operators which are sparse $N\times N$ matrices. $N$ being the total number of vertices of the underlying graph.
For the purpose of modeling a QSW with only pure dephasing, the Lindblad operators are (see section 4.2 of Ref.~\cite{QSWalk} and Refs.~\cite{kendon,QSW_def} for more details on modeling QSW with pure dephasing),
\begin{equation}
        \hat{L}_{k}={\sqrt{|H_{ii}|}}\mbox{  } {\mid\!i\rangle\langle i\!\mid},
        \label{pure-dep}
    \end{equation}
 wherein $H_{ii}= \bra{i} H \ket{i}$ are the diagonal elements in the matrix representation of Hamiltonian operator $H$. {The term ``pure dephasing'' has different significance in different contexts. In some cases, pure dephasing represents fluctuation of energy in eigenbasis set. This fluctuation of energy leads results in incoherence but the population is still conserved~\cite{weiss2012quantum}. In our case, as shown in Ref.~\cite{QSW_def} too, however, it means to represent the fluctuation of the diagonal elements of the Hamiltonian in the site representation due to interaction with the environment. Since the Hamiltonian in this case is not diagonal, the site populations will be modified due to pure dephasing.} For $\omega=1$ it can be shown~\cite{QSWalk} that {, Eq.~\ref{QWS_def} reduces to}
 \begin{equation}
     \frac{d\rho_{ij}(t)}{dt}=\begin{cases}-\rho_{ij}(t),& i \neq j, \\
     0, & i=j.
     \end{cases}
     \label{purw=1}
 \end{equation}
Eq.~(\ref{purw=1}) shows that the off-diagonal terms of the density matrix which represent coherences die exponentially while the population~($\rho_{ii}$) remains constant. So, there is no transport at $\omega=1$ limit for the pure dephasing model. As an aside, and not particularly relevant to exciton transport in FMO, in Ref.~\cite{take} it has been predicted that the QSW model with pure dephasing can be implementated on a quantum computer.\\
The Lindblad operators corresponding to only incoherence are given by
\begin{equation}
        \hat{L}_{k}={\sqrt{|H_{ij}|}} \mbox{  } {\mid\!\!i\rangle\langle j\!\!\mid},
         \label{incoh}
    \end{equation}
with $H_{ij}= \bra{i} H \ket{j}$ being the matrix elements of the Hamiltonian operator $H$ with the condition $i\neq j$. {``Only incoherence'' means fluctuations of the off-diagonal elements of the Hamiltonian in site representation.} For the $\omega=1$ limit, QSW with only incoherent scattering reduces to CRW.\\
On the other hand, when modeling a QSW with both dephasing and incoherence, the Lindblad operators are chosen to be, (see also~\cite{QSW_def} and section 2 of Ref.~\cite{QSWalk})-
\begin{equation}
        \hat{L}_{k}={\sqrt{|H_{ij}|}} \mbox{  } {\mid\!\!i\rangle\langle j\!\!\mid},
         \label{dep-incoh}
    \end{equation}
with $H_{ij}= \bra{i} H \ket{j}$ being the matrix elements of the Hamiltonian operator $H$. The Lindblad operators in Eq.~(\ref{dep-incoh}) represent scattering between all pair of vertices. {``Dephasing with incoherence'' represents fluctuations in all elements of the Hamiltonian in site representation.} Similar to the last case, here as well, for $\omega= 1$, this QSW model reduces to CRW. Furthermore, for all three schemes, at $\omega=0$, the QSW reduces to CTQW.
\section{Modeling exciton transport in FMO complex}
In the photosynthetic process, energy gathered from light by antennae molecules is transmitted across the network of chlorophyll molecules to reaction centers~\cite{Dynamics}. This process is nearly $100\%$ efficient as nearly all the red photons are captured and stored as energy. Experiments have revealed long-lasting quantum coherence in energy transport across a range of  photosynthetic light-harvesting complexes~\cite{Engel2007,E1,E2,E3,E4}. One such light-harvesting complex is the Fenna-Matthews-Olson (FMO) complex from green sulphur bacteria. The FMO complex consists of seven sites called chromophores, through which the excitons propagate via hopping and it has been speculated that this propagation is aided by phase coherence~\cite{Engel2007}.
\subsection{Understanding exciton transport in FMO: the model of Hoyer, et. al.,~\cite{Hoyer_2010}}
To understand how quantum effects are important in the very efficient transfer of excitons at room temperatures, Hoyer, et. al., in Ref.~\cite{Hoyer_2010} used a master equation approach to model exciton transport in FMO by incorporating only incoherence into their model. The plot of total site coherence (Fig.~3(c) of \cite{Hoyer_2010}) at both $77~K$ and $300~K$ indicates the presence of coherence for nearly $500~fs$. Despite this long lived coherence, the plot of power law of mean square displacement (Fig.~3(b) of \cite{Hoyer_2010}) indicates that much of the transport occurs in the sub-diffusive regime. The super-diffusive nature of transport lasts for only $70~fs$, indicating localization time of $70~fs$ at both temperatures of $77~K$ and $300~K$. Hence, Hoyer, et. al., in Ref.~\cite{Hoyer_2010} conclude that coherence in FMO does not yield dynamic speed up unlike that seen in quantum search algorithms, even though quantum coherence may last longer. The main conclusion of  Ref.~\cite{Hoyer_2010} was quantum coherent effects in a biological system like FMO lead to optimized or robust exciton transport, which is $100\%$ efficient rather than any speedup of the transport. Beside, Hoyer, et. al.'s work, there have been many different approaches to exciton transport, see e.g., Refs.~\cite{doi:10.1021/jp109559p, doi:10.1063/1.3155372}. In this work we test this conclusion regarding speedup of the exciton transport via quantum stochastic walks (QSW). To this end, we employ three different approaches for incorporating incoherence into the QSW, the first pure dephasing (see Eq.~(\ref{pure-dep})), the second  with only incoherence  (see Eq.~(\ref{incoh})) and the third dephasing with incoherence(see Eq.~(\ref{dep-incoh})). All three methods are explained in greater detail in the subsequent sections. 
\subsection{Modeling FMO using QSW}
Earlier attempts~\cite{41,43,Hoyer_2010} to model exciton transfer in FMO complex have used non-probability conserving master equations. These equations use loss terms to model absorption of excitons by reaction centers and combine incoherent and coherent transport via a master equation approach. Here, we take another approach of adding an extra vertex and model FMO using probability conserving QSW. This was first done in Ref.~\cite{QSWalk}. This model is versatile and can reproduce pure dephasing transport as well as both dephasing and incoherent scattering, however, it does not have an explicit temperature dependence. Ref.~\cite{QSWalk} also made an incorrect comparison of the QSW model with both incoherence and dephasing to the model of Ref.~\cite{Hoyer_2010} which includes only incoherence. In our work, we correct this and compare QSW with only incoherence to the model of Ref.~\cite{Hoyer_2010} to get a good match of $\omega$ with temperature. Thus, we compare the plots of total site coherence versus time for various values of $\omega$ for QSW with only incoherence and Fig.~3(c) of Ref.~\cite{Hoyer_2010}. We find that $\omega=0.19$ corresponds roughly to a temperature of $77~K$ and $\omega=0.45$ corresponds approximately to a temperature of $300~K$. Using $\omega=0.19$ in the simulation of QSW with only incoherence in FMO, the plot of site population versus time shows similar behavior as Fig.~3(d) of Ref.~\cite{Hoyer_2010} which is at $77~K$. Further, for $\omega=0.45$, the total site coherence versus time shows similar behavior to Fig.~3(c) of Ref.~\cite{Hoyer_2010} which is at $300~K$. First in sub-sub-section III.B.1 we model exciton transfer in FMO using QSW with pure dephasing, then in sub-sub-section III.B.2 only incoherent scattering is considered and finally, in sub-sub-section III.B.3, we make use of both dephasing and incoherent scattering in QSW. The QSW simulations shown in section IV use the \textit{QSWalk} package~\cite{QSWalk,QSWalkdownload} for \textit{Wolfram Mathematica}.
\subsubsection{Modeling exciton transport in FMO via QSW with pure dephasing scattering}
\begin{figure}[t]
    \includegraphics[width=0.4\textwidth]{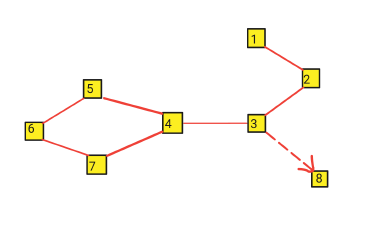}
    \vskip -0.2in
    \caption{Simplified figure of FMO complex. The chromophores are numbered from $1$ to $7$. Vertex $8$ is the sink. Exciton travels from site $6$ and finds it's way to site $3$ the reaction center. The lines between the chromophore sites represent dipolar coupling between them. In fact, there is coupling between every site which is represented by the Hamiltonian given in Eq.~(\ref{hamiltonian})~\cite{ADOLPHS}. But only the couplings above $15 cm^{-1}$ have been shown in Fig.~\ref{FMO}, similar to Ref.~\cite{Hoyer_2010}.}
    \label{FMO}
\end{figure}

FMO has $7$ chromophore sites (see Fig.~\ref{FMO}). The excitation starts at initial site $6$ and gets absorbed at site $3$~\cite{ADOLPHS}, the reaction center. QSW is a probability conserving process, in contrast Hoyer, et. al.'s model~\cite{Hoyer_2010} doesn't conserve probability. To have a one-to-one correspondence between all of our QSW models with the model of Hoyer, et. al., we include a sink to model absorption.  Therefore, in our QSW simulations, an extra site numbered $8$ is added, which acts as a sink, see also Ref.~\cite{QSWalk}. This extra vertex is added using a directed edge and does not take part in the coherent transport via the Hamiltonian. 
The time evolution of the  density matrix for our QSW is given by Eq.~(\ref{QWS_def}).
We use the Hamiltonian obtained by Adolphs and Renger~\cite{ADOLPHS}:
    \begin{equation}
        H=\begin{bmatrix}
        200&-96&5&-4.4&4.7&-12.6&-6.2\\
        -96&320&33.1&6.8&4.5&7.4&-0.3\\
        5&33.1&0&-51.1&0.8&-8.4&7.6\\
        -4.4&6.8&-51.1&110&-76.6&-14.2&-67\\
        4.7&4.5&0.8&-76.6&270&78.3&-0.1\\
        -12.6&7.4&-8.4&-14.2&78.3&420&38.3\\
        -6.2&-0.3&7.6&-67&-0.1&38.3&230\\
                \end{bmatrix}
                \label{hamiltonian}
    \end{equation}
The units of energy are $cm^{-1}$ (we follow the usual spectroscopy convention of expressing energy in terms of the wavelength of photon with that energy, i.e., $1 cm^{-1} \equiv 1.23984\times 10^{-4}eV$). The model uses the above Hamiltonian, padded with zeros to construct an $8 \times 8$ matrix, so as to describe coherent evolution. For the case of QSW model with pure dephasing, we use the set of Lindblad operators, see Eq.~(\ref{dep-incoh}), ${\hat{L}}_{k}=\sqrt{|H_{ii}|} |i\rangle \langle i|$. There is a unique value of $k$ for each vertex pair $i,j$, in this case $i=j$. For this case, the sum in Eq.~(\ref{QWS_def}) extends over all $i$ such that $K=N$. Here, we get CTQW for $\omega=0$ while $\omega=1$ shows no transport for the pure dephasing scattering scheme. An extra Lindblad operator is used for the sink: ${\hat{L}}_{k}=\sqrt{\alpha}|8\rangle\langle 3|$, where $\alpha$ determines the rate of absorption at the sink (here $\alpha=100$ as in Ref.~\cite{QSWalk}). Through these Lindblad operators we incorporate pure dephasing scattering. The initial density matrix is generally taken as $\rho(0)=|6\rangle\langle 6|$, assuming the photon is captured at site $6$.
\begin{figure*}[!]
\subfigure[]{\includegraphics[width=0.32\textwidth]{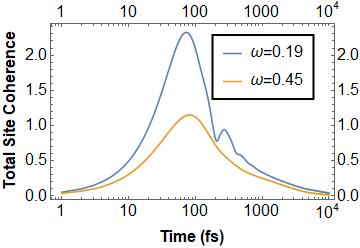}\label{tsc}}
\subfigure[]{\includegraphics[width=0.32\textwidth]{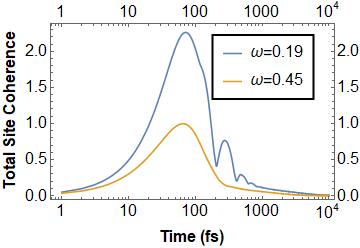} \label{tsc_onlyincoherent}}
\subfigure[]{\includegraphics[width=0.32\textwidth]{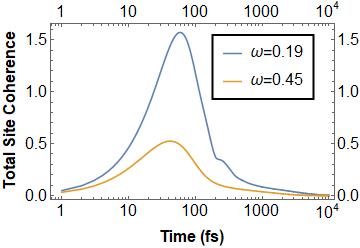} \label{tsc_incoherent}}
\caption{Total site coherence versus time using (a) Pure dephasing (b) Only incoherent scattering. Comparing our data with data from Fig.~3(c) of Ref.~\cite{Hoyer_2010}, we get the equivalent values for $\omega$ at temperatures of $77~K$ and $300~K$. The trend is similar for both our result for $\omega=0.19~(\sim 77~K)$ and Ref.~\cite{Hoyer_2010}, however, there is an offset between the two. It maybe due to the fact that the master equation in Ref.~\cite{Hoyer_2010} differs from ours as Lindblad operators differ. (c) Dephasing and incoherent scattering.}
\end{figure*}
\begin{figure*}[!]
\subfigure[]{\includegraphics[width=.4\textwidth]{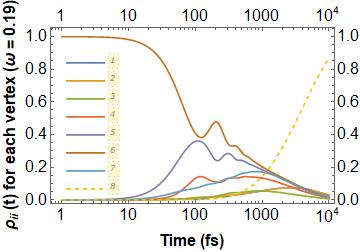}\label{puredeph77Ksitepop}}
\subfigure[]{\includegraphics[width=0.4\textwidth]{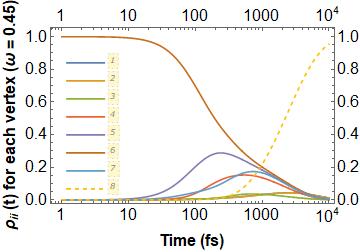} \label{puredeph300Ksitepop}}
\caption{Site population ($\rho_{ii}$, $i=1-8$) versus time for pure dephasing at (a) $\omega=0.19~(\sim 77~K)$ (b) $\omega=0.45~(\sim 300~K)$. Legend indicates population of sites $1-8$.}
\label{d_sitepop}
\end{figure*}
\begin{figure*}[!]
\centering
\subfigure[]{   \includegraphics[width=.4\textwidth]{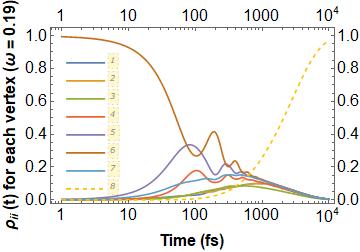}\label{dec77sitepop}}
\subfigure[]{    \includegraphics[width=.4\textwidth]{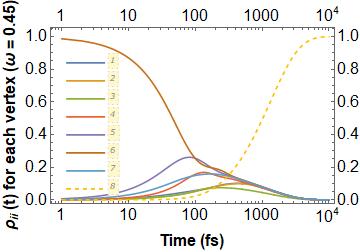}  \label{dec300sitepop}}
\caption{Site population ($\rho_{ii}$, $i=1-8$) versus time for only incoherence at (a) $\omega=0.19~(\sim 77~K)$ (b) $\omega=0.45~(\sim 300~K)$. Legend indicates population of sites $1-8$. Fig.~\ref{dec77sitepop} shows a similar trend as Fig.~3(d) of Ref.~\cite{Hoyer_2010} but again there is an offset between our result and Fig.~3(d) of Ref.~\cite{Hoyer_2010}.}
\label{sitepopincoh}
\end{figure*}
\begin{figure*}[!]
\centering
\subfigure[]{   \includegraphics[width=.4\textwidth]{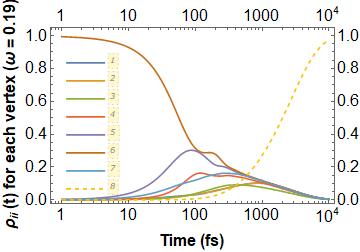}\label{sitepop77Kdephanddec}}
\subfigure[]{    \includegraphics[width=.4\textwidth]{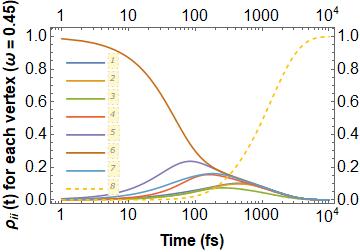}  \label{sitepop300Kdephanddec}}
\caption{Site population ($\rho_{ii}$, $i=1-8$) versus time for dephasing and incoherence at (a) $\omega=0.19~(\sim 77~K)$ (b) $\omega=0.45~(\sim 300~K)$. Legend indicates population of sites $1-8$.}
\label{sitepop}
\end{figure*}

 The pure dephasing model is not new to quantum open systems using master equation approach. In fact one of the first papers to deal with exciton transport in photosynthetic complexes, see Ref.~\cite{rebentrost2009environment}, used a pure dephasing approach.  The justification for using it is that  quantum transport efficiency in a photosynthetic complex can be enhanced by a dynamical interplay of the system Hamiltonian with pure dephasing induced by a fluctuating environment while fully coherent hopping leads to localization.  However, the use of pure dephasing model in quantum stochastic walks per se and the definition of Lindblad operators as in Eq.~(2) where the dephasing rate is fixed by the matrix element of the Hamiltonian, was first given in the paper which introduced quantum stochastic walks, Ref.~\cite{QSW_def}. In this paper it has been shown that a pure dephasing model of quantum stochastic walks does not provide a mapping to a classical random walk. 
\subsubsection{Modeling exciton transport in FMO via QSW with only incoherence}
For QSW using only incoherence, we use the Lindblad operators- defined as in Eq.~(\ref{incoh}), ${\hat{L}}_{k}=\sqrt{|H_{ij}|}|i\rangle\langle j|$ where $i\neq j$. There is a unique value of $k$ for each vertex pair $i,j$. In contrast to the pure dephasing transport scheme, here, the sum in Eq.~(\ref{QWS_def}) extends over all distinct $i,j$ such that $K=N(N-1)$. It is important to note that for this model of QSW, we get CRW ($\omega=1$) and CTQW ($\omega=0$) as two extreme cases. For the sink we use the same Lindblad operator as previous: $\hat{L}_{k}=\sqrt{\alpha}|8\rangle\langle3|$, with $\alpha=100$ determining the absorption rate at the sink.
\subsubsection{Modeling exciton transport in FMO via QSW with both dephasing and incoherent scattering}
Finally, for the case of QSW model with both dephasing and incoherence~\cite{QSW_def} we use the set of Lindblad operators, see Eq.~(\ref{dep-incoh}), ${\hat{L}}_{k}=\sqrt{|H_{ij}|} |i\rangle \langle j|$. Similar to the only incoherent transport scheme, this QSW scheme also results in CRW ($\omega=1$) and CTQW ($\omega=0$) as two extreme cases. For the sink we use the same Lindblad operator as the above two cases: $\hat{L}_{k}=\sqrt{\alpha}|8\rangle\langle3|$, with $\alpha=100$ determining the absorption rate at the sink. %For $\alpha=0$ and $\omega=1$, we find the result for this scheme identical to that for only incoherent scattering scheme i.e., all site populations converge to around 0.15. The interpretation is also the same as the only incoherent scattering scheme i.e., the exciton now has equal probability of being in any of the 7 sites.
We now make a detailed study of exciton transport in FMO complex focusing on total site coherence, site population, mean square displacement and localization time to better understand the exciton dynamics of a FMO complex and compare with the results of Ref.~\cite{Hoyer_2010} especially with regards to the localization time. Since temperature doesn't appear in QSW, we try to make a one to one correspondence between our simulation of the total site coherence (see section IV.A below) with the same simulation in Ref.~\cite{Hoyer_2010}. This gives us the equivalent $\omega$ values for particular temperatures. $\omega=0.19$ corresponds roughly to a temperature of $77~K$ and $\omega=0.45$ corresponds roughly to $300~K$. This equivalence between $\omega$ and temperature is achieved by comparing the plot of total site coherence Fig.~\ref{tsc_onlyincoherent} of this work, with  Fig.~3(c) of Ref.~\cite{Hoyer_2010}. The procedure to calculate the total site coherence is given in the next section.
\section{Results}
\subsection{Total Site Coherence}
Total site coherence is defined as the sum of the absolute value of each off-diagonal element of the density matrix in the site basis. Finite valued off-diagonal elements of a density matrix indicate coherence. Total site coherence is a measure of the coherence present in the FMO complex. The QSW does not have explicit temperature dependence. It has a single parameter $\omega$ which behaves similar to temperature. Thus, to get an approximate temperature dependence for our QSW model via $\omega$, we compare plots of total site coherence versus time (for only incoherence) at different values of $\omega$ with Fig.~3(c) of Ref.~\cite{Hoyer_2010}. We find that $\omega=0.19$ corresponds roughly to a temperature of $77~K$ while $\omega=0.45$ corresponds roughly to $300~K$ (see Fig.~\ref{tsc_onlyincoherent} which depicts total site coherence for exciton transport in QSW with only incoherence). Ref.~\cite{Hoyer_2010} uses a master equation with only incoherence to model exciton transfer in FMO. Hence the plot of total site coherence using QSW with only incoherent scattering has been compared with the respective plot (Fig.~3(c)) of Ref.~\cite{Hoyer_2010} to obtain a rough correspondence between $\omega$ and temperature. In Figs.~\ref{tsc}, \ref{tsc_onlyincoherent} and \ref{tsc_incoherent}, we plot total site coherence for pure dephasing scattering, only incoherent scattering and dephasing with incoherent scattering respectively. One can see that Fig.~\ref{tsc} has the highest peak and the longest tail while Fig.~\ref{tsc_incoherent} has the shortest peak and the least oscillations. On the other hand, Fig.~\ref{tsc_onlyincoherent} has the most oscillations among the three plots.
\begin{figure*}[!]
\subfigure[]{\includegraphics[width=0.32\textwidth]{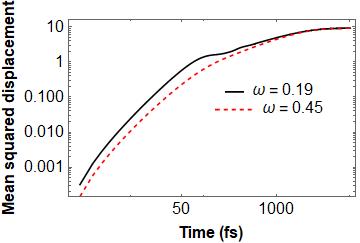}\label{xsqdeph}}
\subfigure[]{\includegraphics[width=0.32\textwidth]{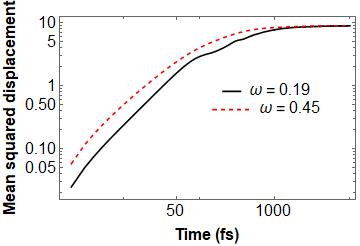}\label{xsqdec}}
\subfigure[]{\includegraphics[width=0.32\textwidth]{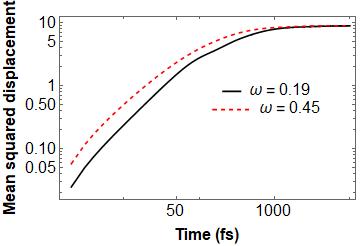}\label{xsqdeph+dec}}
\caption{Log-log plot of mean square displacement versus time for (a) pure dephasing (b) only incoherence and (c) dephasing with incoherence.}
\label{xsq}
\end{figure*}
\begin{figure*}[!]
\subfigure[]{\includegraphics[width=0.32\textwidth]{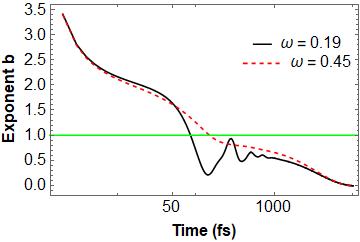}\label{d_powerlaw}}
\subfigure[]{\includegraphics[width=0.32\textwidth]{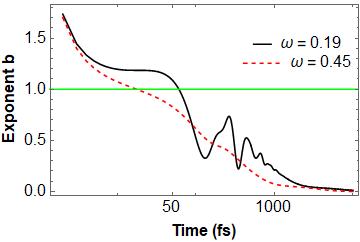}\label{dec_powerlaw}}
\subfigure[]{\includegraphics[width=0.32\textwidth]{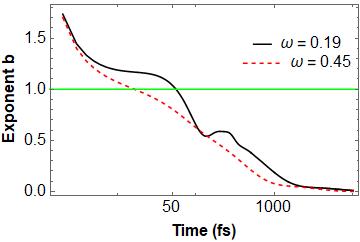}\label{powerlaw}}
\caption{Power law $b$ versus time for (a) pure dephasing (b) only incoherence and (c) dephasing with incoherence.}
\label{expb}
\end{figure*}
\subsection{Site Population}
%Now, we try to find what happens if we take the case $\alpha=0$ and $\omega=1$, i.e., when the absorption rate of the sink is zero and the temperature is extremely high. We end up with the result that all the 7 site populations converge to around 0.15 for this scheme while site population for site 8 remains at zero. This means that the exciton, while starting from site 6, now has equal probability of being in any of the 7 sites.
Site population of the $i^{th}$ site in the FMO complex is defined as ${ii}^{th}$ element of the density matrix $\rho$, denoted as $\rho_{ii}^{th}$. Site population of the $i^{th}$ site represents the probability of finding the exciton at that site. For example, if we had to calculate the site population at site $5$ (say), we just have to extract element $\{5,5\}$ of the density matrix. Similarly, we can extract the site population of all sites. Initially the exciton is at site $6$. Evolving the density matrix using Eq.~(\ref{QWS_def}), the population of each site can be calculated. Site population with pure dephasing scattering has been plotted in Fig.~\ref{d_sitepop} at both $\omega=0.19~(\sim77~K)$ and $\omega=0.45~(\sim300~K)$. We observe oscillating site populations in Fig.~\ref{puredeph77Ksitepop} which indicate persistent coherence. With increase in $\omega$, the oscillations disappear since transport becomes more incoherent as can be seen in Fig.~\ref{puredeph300Ksitepop}. Now, we try to find what happens if we take the case $\alpha=0$ and $\omega=1$, i.e., when the absorption rate of the sink is zero and the temperature is extremely high. We see that all the site populations except the initial site remain at zero while the initial site i.e., site $6$ has a population of $1$. This implies that there is no transport of exciton for this case. For only incoherent transport scheme, site population versus time can be seen in Fig.~\ref{sitepopincoh}. As mentioned above, here as well in Fig.~\ref{dec77sitepop}, one can see oscillations indicating persistent coherence (highest for this scheme) which stabilise for $\omega=0.45~(300~K)$. Fig.~\ref{dec77sitepop} shows a similar trend as in Fig.~3(d) of Ref.~\cite{Hoyer_2010} with the only difference being an offset in the data. Now, once again, when we substitute $\alpha=0$ and $\omega=1$, we end up with the result that all the $7$ site populations converge to around $0.15$ for this scheme while site population for site $8$ remains at zero. This means that the exciton, while starting from site $6$, now has equal probability of being in any of the $7$ sites. We do the same for dephasing with incoherence transport. The site population versus time has been plotted in Fig.~\ref{sitepop}. Once again, we find oscillations (least for this case) for $\omega=0.19$, see Fig.~\ref{sitepop77Kdephanddec} which disappear at $\omega=0.45$. Again when we substitute $\alpha=0$ and $\omega=1$, we find the result for this scheme identical to that for only incoherent scattering scheme, i.e., all site populations converge to around $0.15$. The interpretation is also the same as the only incoherent transport i.e., the exciton now has equal probability of being in any of the $7$ sites.
%Site population versus time for QSW with only incoherence at both $77~K$ (i.e., $\omega=0.19$) and $300~K$ (i.e., $\omega=0.45$) has been plotted in Fig.~\ref{sitepopincoh}. Fig.~\ref{dec77sitepop} shows a similar trend as in Fig.~3(d) of Ref.~\cite{Hoyer_2010} with the only difference being an offset in the data. Site population versus time for QSW with only incoherence is given in Fig.~\ref{sitepopincoh}. Now, we try to find what happens if we take the case $\alpha=0$ and $\omega=1$, i.e., when the absorption rate of the sink is zero and the temperature is extremely high. We end up with the result that all the 7 site populations converge to around 0.15 for this scheme while site population for site 8 remains at zero. This means that the exciton, while starting from site 6, now has equal probability of being in any of the 7 sites.
%that with pure dephasing is given in Fig.~\ref{d_sitepop}  Similar to the previous subsection, we substitute $\alpha=0$ and $\omega=1$. In contrast to the scheme with only incoherent scattering, in this scheme, all the site populations except the initial site remain at zero while the initial site i.e., site 6 has a site population of 1. This implies that there is no transport of exciton for this case.
%and that with both incoherence and dephasing is given in Fig.~\ref{sitepop}. We see, in all three cases, that with increase in temperature ($\omega$) the oscillations die down faster with time.
\subsection{Mean Square Displacement}
Mean square displacement is defined as~\cite{Hoyer_2010}
\begin{equation}
    \langle x^{2}\rangle=\frac{Tr(\rho x^{2})}{Tr(\rho)},
    \label{meansqdisp}
\end{equation}
where, $x$ is the displacement from the initial site with $\rho$ as the density matrix. The mean square displacement depicts how fast the exciton moves away from the initial site. The mean square displacement versus time has been plotted in Fig.~\ref{xsqdeph} for pure dephasing, Fig.~\ref{xsqdec} for only incoherence and Fig.~\ref{xsqdeph+dec} for dephasing with incoherence. We assume a power law relation (see Eq.~(\ref{msqdisp})) between the mean square displacement $\langle x^2\rangle$ and time $t$ since we wish to test the extent of quantum speed up. This is so because $b=2$ in Eq.~(\ref{msqdisp}) corresponds to ideal quantum speed up while $b=1$ corresponds to the limit of diffusive transport~\cite{Hoyer_2010}.
\begin{eqnarray}
\langle x^{2}\rangle &=& t^{b}, \mbox{ and taking logarithm on both sides,}\nonumber\\
\mbox{ we get-  } \log \langle x^{2}\rangle &=& b \log t.
\label{msqdisp}
\end{eqnarray}
where $t$ denotes time. Exponent `$b$' versus time has been plotted in Fig.~\ref{expb}, which has been obtained from slope of log-log plot of mean square displacement (slope of Fig.~\ref{xsq}). For $b>1$ the transport is called super diffusive and $b<1$ corresponds to sub-diffusive transport. These definitions are with respect to classical random walk (CRW) which follows diffusive transport at $b=1$. In Fig.~3(b) of Ref.~\cite{Hoyer_2010} the plots of exponent $b$ versus time show the transition from super diffusive to sub-diffusive transport at $70~fs$ for both $77~K$ as well as $300~K$. These exponent $b$ versus time plots for QSW with pure dephasing (Fig.~\ref{d_powerlaw}), only incoherence (Fig.~\ref{dec_powerlaw}) and that with both dephasing and incoherence (Fig.~\ref{powerlaw}) show that the time for this transition is different at different $\omega$ (corresponding to different temperature). More about this result has been explained in the next sub-section on localization time. Further, the \textit{Mathematica} code and method of calculation for mean square displacement, exponent $b$ and localization time ($t_{loc}$) has been provided in Appendix A.
\begin{table*}[!]
\caption{Localization time comparison for Model of Hoyer, et. al.,Ref.~\cite{Hoyer_2010}, QSW with pure dephasing (Eq.~(\ref{pure-dep})), QSW with only incoherence (Eq.~(\ref{incoh})) and dephasing with incoherence (Eq.~(\ref{dep-incoh})) in exciton transfer through FMO.}
\begin{tabular}{{|c|c|c|c|c|}}
\hline
\textbf{} &
  \begin{tabular}[c]{@{}c@{}}Model of Hoyer,\\ et. al.,Ref.~\cite{Hoyer_2010}\end{tabular} &
  \begin{tabular}[c]{@{}c@{}}QSW with pure\\ dephasing (Eq.~\ref{pure-dep})\end{tabular} &
  \begin{tabular}[c]{@{}c@{}}QSW with only\\incoherence (Eq.~\ref{incoh})\end{tabular} &
  \begin{tabular}[c]{@{}c@{}}QSW with dephasing\\and incoherence (Eq.~\ref{dep-incoh})\end{tabular} \\
  \hline
\begin{tabular}[c]{@{}c@{}}$t_{loc} $ at $\omega=0.19~(\sim77~K)$\\ (in femto-secs)\end{tabular} &
  70 &
  85.41 &
  60.91 &
  55.97 \\
  \hline
\begin{tabular}[c]{@{}c@{}}$t_{loc} $ at $\omega=0.45~(\sim300~K)$\\ (in femto-secs)\end{tabular} &
  70 &
  145.86 &
  17.15 &
  16.21\\
\hline
\end{tabular}
\label{tlocdata}
\end{table*}
\subsection{Localization time}
\begin{figure*}[!]
\subfigure[]{\includegraphics[width=0.32\textwidth]{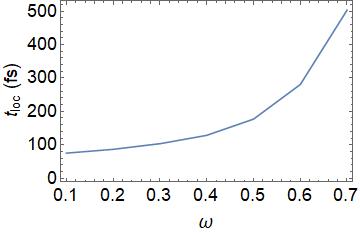}\label{tlocvswdeph}}
\subfigure[]{\includegraphics[width=0.32\textwidth]{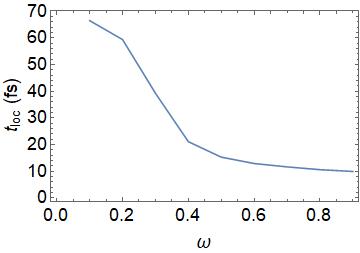}\label{tlocvswdec}}
\subfigure[]{\includegraphics[width=0.32\textwidth]{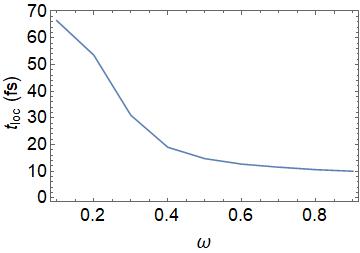}\label{tlocvswdephanddec}}
\caption{Localization time versus $\omega$ for (a) pure dephasing (b) only incoherence and (c) dephasing with incoherence.}
\label{tlocvsw}
\end{figure*}

The localization time ($t_{loc}$)~\cite{Hoyer_2010} is defined as the time at which the transition of power law $b$ occurs from super diffusive to the sub-diffusive regime, i.e., $b$ goes below $1$. $t_{loc}$ values have been given in Table~\ref{tlocdata}. We see that for only incoherence and dephasing with incoherent transport, there is a slowdown instead of a speedup. On the other hand, for the QSW model with pure dephasing there is indeed a speed up at $\omega=0.45~(300~K)$ as $t_{loc}$ increases to $145.86~fs$ as compared to $\omega=0.19~(77~K)$ where it is $85~fs$. These respective trends are still valid if we change the initial state of the exciton to be at site $1$ instead of site $6$. Now, for the only incoherent transport, we observe a slowdown since $t_{loc}$ reduces from $60.91~fs$ at $\omega=0.19$ to $17.15~fs$ at $\omega=0.45$. However, if we look closely at Fig.3(b) in Ref.~\cite{Hoyer_2010}, we can see that the $t_{loc}$, although almost same, is somewhat increased at $300~K$ than it was at $77~K$. This fact is quite intriguing as although both our work and Ref.~\cite{Hoyer_2010} makes use of only incoherence model, due to the difference in Lindblad operators occuring in the respective master equations, the results differ significantly.\\
In case of QSW model with both dephasing and incoherence there is also a slow down instead of speed up, $t_{loc}$ reduces from $55.97~fs$ at $\omega=0.19~(77~K)$ to $16.21~fs$ at $\omega=0.45~(300~K)$. An explanation for these findings on localization time has been provided in Appendix B and Appendix C by comparing site populations. One can clearly see in Fig.~\ref{sitewisepopdec} (Appendix C where we compare site populations for only incoherent transport), that the exciton population at each site invariably reaches a peak at $\omega=0.45$ earlier than that at $\omega=0.19$. Thus, time for $\rho_{ii}$ to peak at $\omega=0.45$ is always less than the time for $\rho_{ii}$ to peak at $\omega=0.19$ for QSW model with only incoherent scattering. One can also see in Fig.~\ref{sitewisepopdec} (green curve), the plot of Hoyer, et. al., for each site and one can see that it is always later to peak than for QSW model with only incoherence at $\omega=0.19$.\\
In contrast in Fig.~\ref{sitewisepopdeph} (Appendix B where we compare site populations for pure dephasing transport), one can clearly see that the exciton population at each site invariably reaches a peak later at $\omega=0.45$ than at $\omega=0.19$. Thus, time for $\rho_{ii}$ to reach a peak at $\omega=0.19$ is always less than time for $\rho_{ii}$ to reach a peak at $\omega=0.45$ for QSW model with pure dephasing. Similar to only incoherent scattering case, for both dephasing and incoherence, it can be seen from Fig.~\ref{sitewisepopdeph+dec} that $\rho_{ii}$ always reaches a peak earlier for $\omega=0.45$ than it does for $\omega=0.19$. Now, in order to verify if the trend described above for the localization time for all the three schemes is followed for all values of the parameter $\omega$, we plot the behaviour of localization time with changing $\omega$ in Fig.~\ref{tlocvsw}. We see that for dephasing with incoherence (Fig.~\ref{tlocvswdephanddec}) and only incoherent scattering (Fig.~\ref{tlocvswdec}) transport, localization time decreases monotonically with increasing $\omega$, implying a slowdown. On the other hand, we see a monotonous increase in the localization time as $\omega$ increases for the pure dephasing transport (Fig.~\ref{tlocvswdeph}), thus implying a speedup in exciton transport.
\begin{table*}[!]
  \caption{Comparison of three models (Model of Hoyer, et. al.,~\cite{Hoyer_2010},  QSW with pure dephasing, QSW with both incoherence and dephasing and QSW with only incoherence) for exciton transfer in FMO. } 
\begin{tabularx}{\textwidth}{|l|X|X|X|X|}
\hline
&Model of Hoyer, et. al.,~\cite{Hoyer_2010}& QSW model with pure dephasing& QSW with only incoherence & QSW model with dephasing and incoherence \\
\hline\hline
 Total Site Coherence& Similar to case of QSW with only incoherence (Fig.~3(c) in \cite{Hoyer_2010}).& Decreases with rise in $\omega$ (Fig.~\ref{tsc}). & Decreases with rise in $\omega$ (Fig.~\ref{tsc_onlyincoherent}). & Decreases with rise in $\omega$ (Fig.~\ref{tsc_incoherent}).\\
\hline
Site Population&Similar to QSW using only incoherence (Fig.~3(d) in \cite{Hoyer_2010}).& Oscillations are more prominent for dephasing but decrease with rise in $\omega$ (Fig.~\ref{d_sitepop}). & Oscillations in site population vanish with increasing $\omega$ (Fig.~\ref{sitepopincoh}).& Oscillations in site population vanish with increasing $\omega$ (Fig.~\ref{sitepop}). \\
\hline
Power law $b$ & Super diffusive to sub-diffusive transition occurs at $70~fs$ for both $77~K$ and $300~K$ (Fig.~3(b) in \cite{Hoyer_2010}). & Super-diffusive to sub-diffusive transition occurs at $85.41 fs$ for $\omega=0.19$ and $145.86 fs$ for $\omega=0.45$ (Fig.~\ref{d_powerlaw}). & Super-diffusive to sub-diffusive transition occurs at $60.91 fs$ for $\omega=0.19$ and $17.15 fs$ for $\omega=0.45$ (Fig.~\ref{dec_powerlaw}). & Super-diffusive to sub-diffusive transition occurs at $55.97 fs$ for $\omega=0.19$ and $16.21 fs$ for $\omega=0.45$ (Fig.~\ref{powerlaw}).\\
\hline
\hline
\end{tabularx}
\label{modelcomp}
\end{table*}
\section{Discussion and Conclusion}
Table~\ref{tlocdata} compares the localization time for the three transport schemes. It is evident that no speedup is observed in case of Ref.~\cite{Hoyer_2010}. We see for only incoherent transport and also for a transport incorporating both incoherence and dephasing, a significant slowdown. On the other hand, the localization time for pure dephasing scheme increases at $\omega=0.45$ as compared to $\omega=0.19.$ This is in line with the quantum Goldilocks effect~\cite{Jim,Goldilocks} which predicted increase in speed of exciton transport at a temperature nearly equal to the room temperature. Neither the model of Ref.~\cite{Hoyer_2010} nor the QSW model with both dephasing and incoherence and that with only incoherence can explain this effect as the localization time is same for both $77~K$ and $300~K$, in case of Ref.~\cite{Hoyer_2010} while in cases of the QSW model with both dephasing and incoherence and the one with only incoherent scattering, there is a slow down instead of speed up, rendering any quantum effect meaningless, which is in line with recent works on photosynthetic energy transfer which debunk any speedup due to quantum coherence~\cite{duan}. This result has major implication for studies in exciton transport through FMO. It means a QSW model with pure dephasing is best able to explain not only the robust transport of exciton but also the quantum advantage which delivers the necessary speed up to exciton transport process. This is in line with earlier study which predicted~\cite{Goldilocks} maximum efficiency of exciton transport process in FMO around room temperatures than very low temperatures like $77~K$. In FMO complex, pure dephasing model ignores exciton relaxation and spatial correlations in the environment~\cite{rebentrost2009environment}.

\begin{figure*}[!]
\centering
\subfigure[]{   \includegraphics[width=0.32\textwidth]{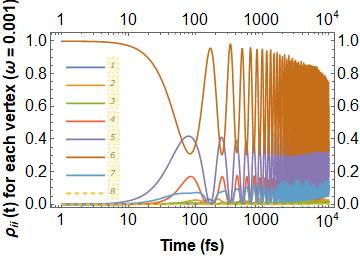}\label{0.001deph}}
\subfigure[]{    \includegraphics[width=0.32\textwidth]{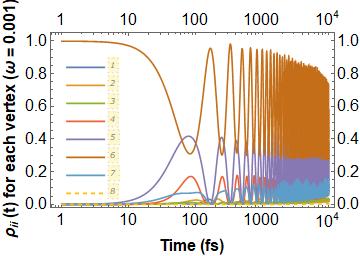}  \label{0.001dec}}
\subfigure[]{    \includegraphics[width=0.32\textwidth]{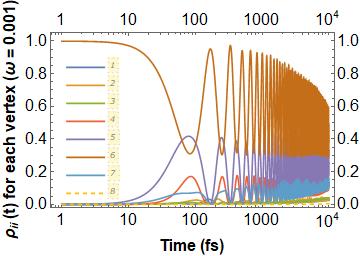}  \label{0.001dephanddec}}
\caption{Plot of site population versus time for $\omega=0.001$ showing persisting oscillations and the exciton coming back to initial site i.e., site 6 repeatedly for QSW model with (a) pure dephasing, (b) only incoherence and (c) dephasing with incoherence.}
\label{w_pop}
\end{figure*}

To check that our model of exciton transfer using QSW is in line with earlier studies, we check the transport process at very high coherence which corresponds to near absolute zero temperature. It has been shown in Ref.~\cite{Goldilocks} that an optimum amount of coherence is required for maximum efficiency of any quantum transport process. This has been called the quantum Goldilocks effect. If the environment is too cold, i.e., $\omega\rightarrow 0$ or  fully  coherent exciton transport, the exciton will wander aimlessly without getting anywhere. In this case the exciton will behave like a wave but will not be able to propagate due to destructive interference. We have shown this effect by putting $\omega=0.001$ (highly coherent) and looking at the site population. The plot of site population (see Fig.~\ref{w_pop}) shows that the exciton keeps coming back to initial site, that is site $6$. Even after $10^4$ femto-seconds the exciton can be found with a  very high probability at site $6$. This shows there is essentially no transport at very low temperatures when transport is fully coherent.
%\begin{figure}[!]
%\centering
    %\includegraphics[width=0.45\textwidth]{w_pop.pdf}
    %\caption{Plot of site population versus time for $\omega=0.001$ showing persisting  oscillations and the exciton coming back to initial site i.e. site 6 repeatedly for QSW model with pure dephasing.}
    %\label{w_pop}
%\end{figure}
In Table~\ref{modelcomp} we compare the three models of exciton transfer in FMO complex as regards the other quantities like total site coherence, site population and exponent in the power law. Having seen that the QSW model with pure dephasing is the closest to describing exciton transport with quantum effects, we see that the total site coherence has larger tail for pure dephasing as compared to other transport schemes. Oscillations in site population in both QSW model with only incoherence and in Hoyer, et. al.'s model mirror each other, the only difference being an offset, while for QSW model with both dephasing and incoherence there is a marked difference with oscillations almost disappearing. The QSW model with pure dephasing also has some oscillations in it, albeit less than that for model using only incoherent scattering. Finally, exponent $b$ in the power law again matches the localization time results seen in Table~\ref{tlocdata}. 

To conclude, quantum stochastic walk is a powerful tool to model exciton transport in FMO complex. QSW has a single parameter $\omega$ that controls how classical effects emerge in the model. With increase in $\omega$, transport becomes more incoherent. The temperature dependence enters the model via $\omega$. This seems intuitive, as with increase in thermal fluctuations the amount of coherence should decrease. Therefore, we compare the plots of total site coherence versus time for different values of $\omega$ for QSW with only incoherence with Fig.~3(c) of Ref.~\cite{Hoyer_2010} to get the corresponding values of $\omega$. Then we model FMO with QSW with only incoherent, pure dephasing as well as both dephasing and incoherent scattering. We find that the QSW model with pure dephasing scattering has increased localization time at $300~K$ as compared to $77~K$ in line with the quantum Goldilocks effect. Neither of the other three models, i.e., QSW model with both dephasing and incoherence, that with only incoherent scattering and the model of Hoyer, et. al., was able to explain this effect as the localization time is same for both $77~K$ and $300~K$ for Ref~\cite{Hoyer_2010} and the other two experience slowdown. QSW model with pure dephasing scattering gives speedup at $\omega=0.45~(300~K)$ as compared to $\omega=0.19~(77~K)$, while both QSW with only incoherence and that with both dephasing and incoherent scattering give slowdowns at $\omega=0.45$. Future works can include studying the transport efficiency in FMO using different initial states, like superposed states, e.g., $\frac{1}{\sqrt{2}}(|{1}\rangle+|{6}\rangle)$ as was done in Ref.~\cite{Zhu2011}). It will also be interesting to study the exciton transfer dynamics via QSW when said entangled initial states (see Ref.~\cite{EntangFMO}) are present. 
\section{Conflicts of interest}
There are no conflicts of interest to declare.
 \section{Acknowledgements}
 PKS would like to thank the National Institute of Science Education and Research, HBNI, Jatni 752050, India for providing hospitality and Dept. of Science and Technology (DST) for the INSPIRE fellowship. CB would like to thank Science and Engineering Research Board (SERB) for funding under MATRICS grant ``Nash equilibrium versus Pareto optimality in N-Player games" (MTR/2018/000070).
 \section{Appendix}
 \subsection{ Mathematica codes for exponent \textit{b} and localization time}
  The \textit{Mathematica} code for plotting mean square displacement and power law $b$ is given here. This has been used to generate Fig.~\ref{xsqdec} and Fig.~\ref{dec_powerlaw}, which is for the case of QSW model with only incoherence. For the case of QSW model with both dephasing and incoherence, we use the Hamiltonian $H$ (defined in the code) instead of using the lists (i.e., we just define the LKSet to be LindbladSet[H]). On the other hand, for pure dephasing model, we just use the diagonal entries of the Hamiltonian to obtain the Lindblad operators. This code makes use of \textit{QSWalk}~\cite{QSWalk,QSWalkdownload} package. For the mean square displacement, we first need to have some notion of distance for the sites of FMO. We refer to Fig.~\ref{FMO} of the FMO complex. The lines between the chromophore sites represent dipolar coupling between them. In fact, there is coupling between every site which is represented by the Hamiltonian given in Eq.~(\ref{hamiltonian})~\cite{ADOLPHS}. But only the couplings above $15 cm^{-1}$ have been shown in Fig.~\ref{FMO}~\cite{Hoyer_2010}. Since the magnitude of coupling less than $15 cm^{-1}$ is pretty low, as also was done in Ref.~\cite{Hoyer_2010}, we choose to ignore these couplings, i.e., these sites are effectively decoupled for the purpose of calculating the effective paths from initial site to reaction center, implying the exciton to be following the path according to Fig.~\ref{FMO}. We can then assign a position to each site accordingly. This approach has also been done in Ref.~\cite{Hoyer_2010} (Fig.~1(b) of \cite{Hoyer_2010}). Since the initial excitation is at site $6$, it is defined to be the origin of FMO. Then according to Fig.~\ref{FMO}, we define the sites $5$ and $7$ to be at a distance of $1$ unit away from site $6$. Similarly, site $4$ to be $2$ units, $3$ being $3$ units, $2$ being $4$ units and $1$ to be $5$ units of distance away from $6$. Since site $8$ is the sink which was added for probability conservation, we assign site $8$ to be $3$ units away from site $6$. With this information, we define the operator $x$ as shown in the code. This operator is an $8\times8$ matrix which has eigenvectors as the sites represented as column vectors and eigenvalues being their respective distance from site $6$, as defined above. We know that expectation value of any operator $\hat{A}$ is given by $\frac{\Tr({\rho} \hat{A})}{\Tr({\rho})}$, ${\rho}$ being the density operator. So the mean square displacement is given by:
 \begin{equation}
    \langle x^{2}\rangle=\frac{\Tr(\rho x^{2})}{\Tr(\rho)}
\end{equation}
where
\begin{equation*}
x=\begin{bmatrix}
        5 & 0 & 0 & 0 & 0 & 0 & 0 & 0\\
        0 & 4 & 0 & 0 & 0 & 0 & 0 & 0\\
        0 & 0 & 3 & 0 & 0 & 0 & 0 & 0\\
        0 & 0 & 0 & 2 & 0 & 0 & 0 & 0\\
        0 & 0 & 0 & 0 & 1 & 0 & 0 & 0\\
        0 & 0 & 0 & 0 & 0 & 0 & 0 & 0\\
        0 & 0 & 0 & 0 & 0 & 0 & 1 & 0\\
        0 & 0 & 0 & 0 & 0 & 0 & 0 & 3\\
                \end{bmatrix}
                \label{x}
\end{equation*}
We have implemented the above equation in the code to find mean square displacement. The slope of the log-log plot of mean square displacement will give the power law $b$, as explained in the section called mean square displacement. The intersection of the power law b plot with the green line in Fig.~\ref{expb} will give us the localization time.
 \begin{widetext}
 {\scriptsize
 \begin{lstlisting}
<< QSWalk` 
energyUnit =2 Pi Quantity["ReducedPlanckConstant"]
Quantity["SpeedOfLight"]/Quantity["Centimeters"]
actionUnit = Quantity["ReducedPlanckConstant"]
timeUnit = UnitConvert[actionUnit/energyUnit, "Femtoseconds"]
N[timeUnit]
H0 = ( {
    {200, -96, 5, -4.4, 4.7, -12.6, -6.2},
    {-96, 320, 33.1, 6.8, 4.5, 7.4, -0.3},
    {5, 33.1, 0, -51.1, 0.8, -8.4, 7.6},
    {-4.4, 6.8, -51.1, 110, -76.6, -14.2, -67},
    {4.7, 4.5, 0.8, -76.6, 270, 78.3, -0.1},
    {-12.6, 7.4, -8.4, -14.2, 78.3, 420, 38.3},
    {-6.2, -0.3, 7.6, -67, -0.1, 38.3, 230}
   } );
x = SparseArray[{{1, 1} -> 5, {2, 2} -> 4 , {3, 3} -> 3, {4, 4} -> 2, {5, 5} -> 1, {6, 6} -> 0, {7, 7} -> 1,
{8, 8} -> 3}, {8, 8}]
H = SparseArray[ArrayRules[H0], {8, 8}];
Clear[rho0];
rho0 = SparseArray[{{6, 6} -> 1}, {8, 8}];
list1 = LindbladSet[H];
list2 = Table[SparseArray[{{i, i} -> Sqrt[H0[[i, i]]]}, {8, 8}], {i, 7}];
LK0 = DeleteCases[list1, Alternatives @@ list2];
LkSet = Append[LK0, SparseArray[{{8, 3} -> 10.}, {8, 8}]];
dt = Quantity["Femtoseconds"]/timeUnit //N
omega = 0.19;
Clear[rho];
qsw[rho_] = QuantumStochasticWalk[H,LkSet,omega,rho,dt]
n = 10000; tfs = Range[0, n];
pt = Chop@NestList[qsw, rho0, n];
a = List[];
slope = List[];
k = 1;
While[k < 10000,
p = pt[[k]]; 
x2 = x.x; m = x2.p;
value = Tr[m]/Tr[p];
 a = Append[a, value]; k++]
p1 = ListLogLogPlot[a, Joined -> True, PlotStyle -> Black, PlotRange -> All,
Frame -> True, FrameLabel -> {"Time(fs)", "Mean squared displacement", RotateLabel -> True];
s = 1;
While[s < 9999,
sl = (Log[a[[s + 1]]] - Log[a[[s]]])/(Log[s + 1] - Log[s]);
slope = Append[slope, sl]; s++]
pl1 = ListLogLinearPlot[slope, PlotRange -> All, Joined -> True, PlotStyle -> Black,
PlotLegends -> {"\[Omega]=0.19}"}, Frame -> True,FrameLabel -> {"time(fs)", "Exponent b", RotateLabel -> True];

omega = 0.45;
Clear[rho]; qsw[rho_] = QuantumStochasticWalk[H, LkSet, omega, rho, dt]
n = 10000;
tfs = Range[0, n];
pt = Chop@NestList[qsw, rho0, n];
a = List[];
slope = List[];
k = 1;
While[k < 10000,
 p = pt[[k]];
 x2 = x.x;
 m = x2.p;
 value = Tr[m]/Tr[p];
 a = Append[a, value]; k++]
s = 1;
While[s < 9999,
 sl = (Log[a[[s + 1]]] - Log[a[[s]]])/(Log[s+ 1] - Log[s]);
 slope = Append[slope, sl]; s++]
p2 = ListLogLogPlot[a, Joined -> True, PlotStyle -> {Red, Dashed}, 
   PlotRange -> All, PlotRange -> All, Frame -> True, 
   FrameLabel -> {"Time(fs)", 
     "Mean squared displacement"}, RotateLabel -> True];
pl2 = ListLogLinearPlot[slope, PlotRange -> All, Joined -> True, 
   PlotStyle -> {Red, Dashed}, PlotLegends -> {"\[Omega]=0.45"}, 
   AxesLabel -> {"Time(fs)", "Exponent b"}];
Show[p1, p2]
pl3 = Plot[1, {x, 0, 1000}, PlotStyle -> Green, Frame -> True, 
FrameLabel -> {"time(fs)", "Exponent b"}, RotateLabel -> True]
Show[pl1, pl2, pl3]
 \end{lstlisting}}
 \begin{figure*}[!]
\subfigure[]{    \includegraphics[width=.24\textwidth]{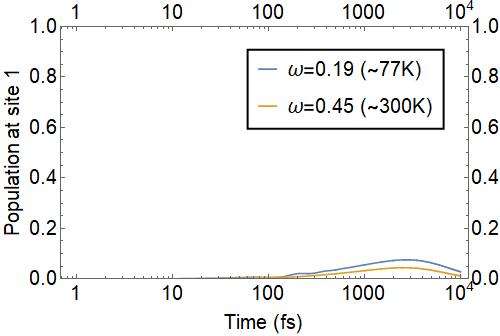}}
    %\label{}
\subfigure[]{    \includegraphics[width=.24\textwidth]{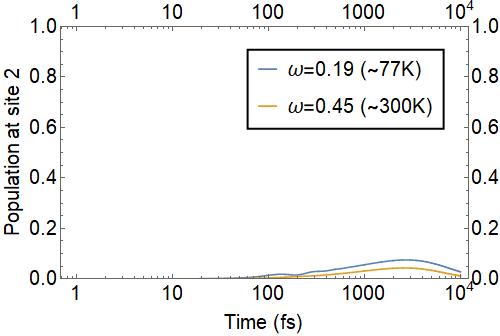}}
    %\label{}
\subfigure[]{    \includegraphics[width=.24\textwidth]{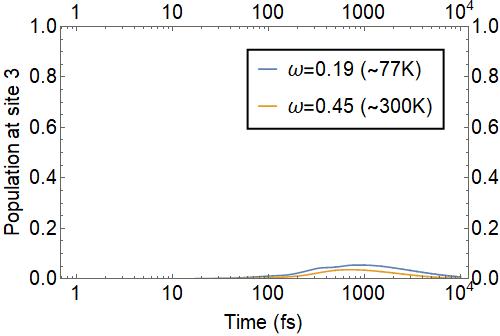}}
    %\label{}
\subfigure[]{    \includegraphics[width=.24\textwidth]{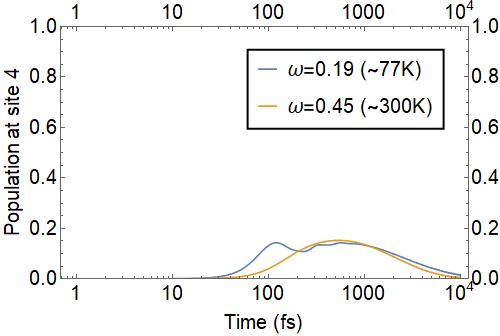}}
    %\label{}
\subfigure[]{    \includegraphics[width=.24\textwidth]{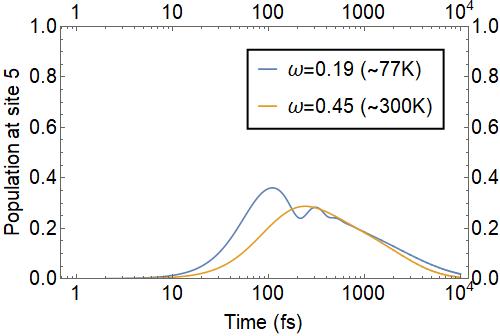}}
    %\label{}
\subfigure[]{    \includegraphics[width=.24\textwidth]{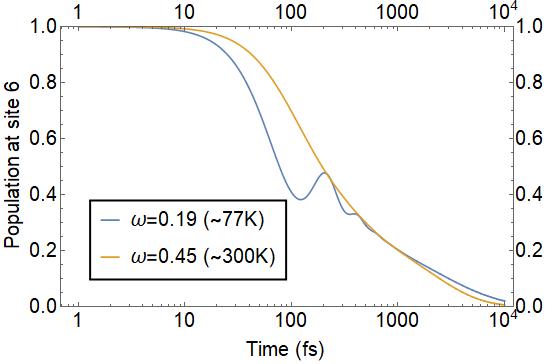}}
    %\label{}
\subfigure[]{    \includegraphics[width=.24\textwidth]{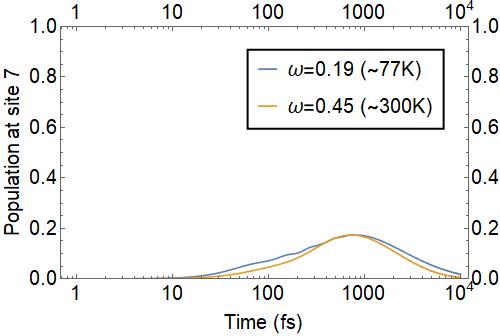}}
    %\label{}
\caption{Site population versus time for various sites at $\omega=0.19$ and $\omega=0.45$ for pure dephasing model.}
\label{sitewisepopdeph}
\end{figure*}
 \begin{figure*}[!]
\subfigure[]{    \includegraphics[width=.32\textwidth]{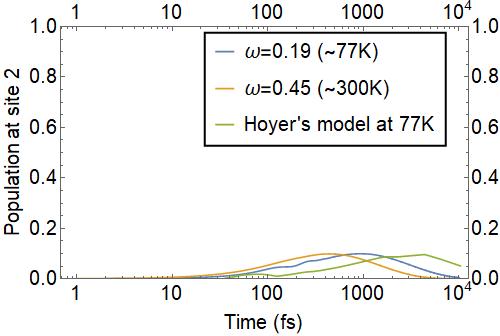}}
    %\label{}
\subfigure[]{    \includegraphics[width=.32\textwidth]{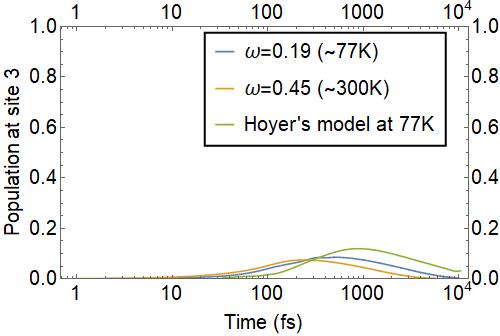}}
    %\label{}
\subfigure[]{    \includegraphics[width=.32\textwidth]{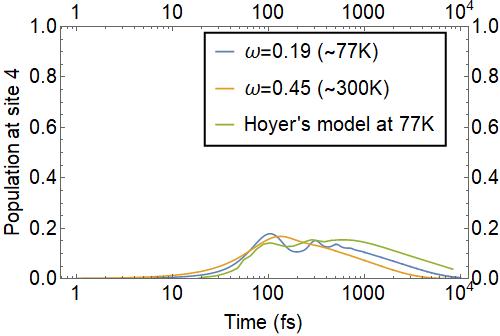}}
    %\label{}
\subfigure[]{    \includegraphics[width=.32\textwidth]{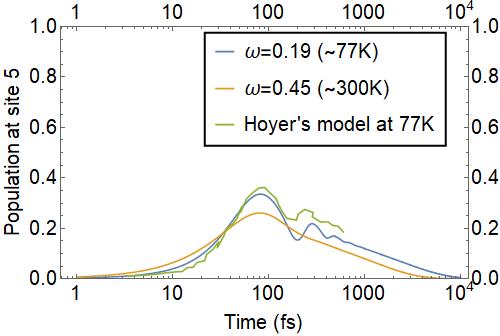}}
    %\label{}
\subfigure[]{    \includegraphics[width=.32\textwidth]{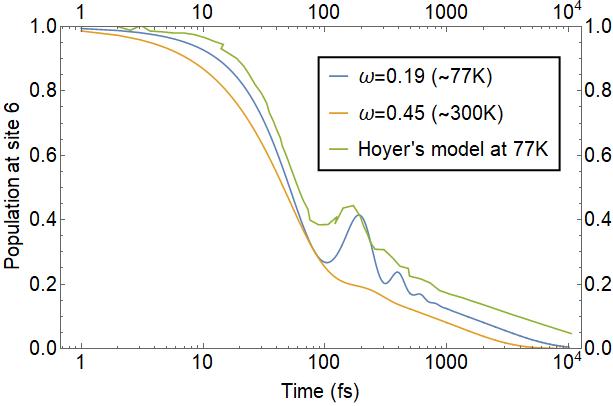}}
    %\label{}
\subfigure[]{    \includegraphics[width=.32\textwidth]{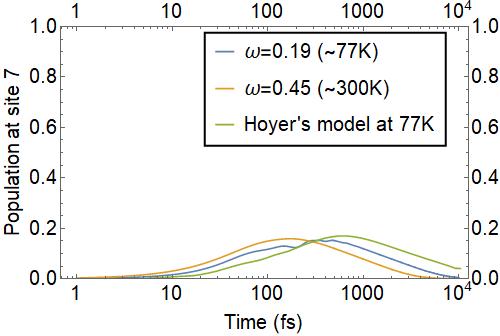}}
    %\label{}
\caption{Site population versus time for various sites at $\omega=0.19$ and $\omega=0.45$. Hoyer's model has been compared with QSW with only incoherence model.}
\label{sitewisepopdec}
\end{figure*}
\begin{figure*}[!]
\subfigure[]{    \includegraphics[width=.24\textwidth]{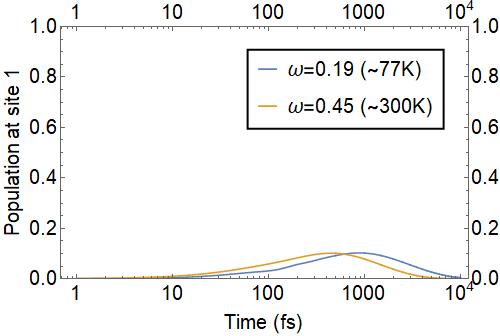}}
    %\label{}
\subfigure[]{    \includegraphics[width=.24\textwidth]{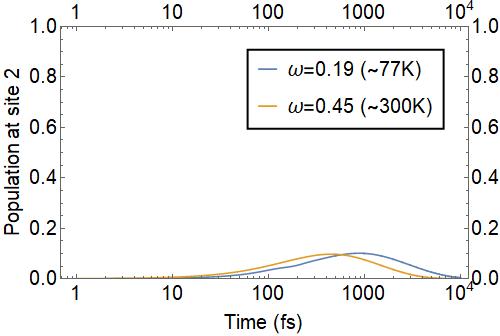}}
    %\label{}
\subfigure[]{    \includegraphics[width=.24\textwidth]{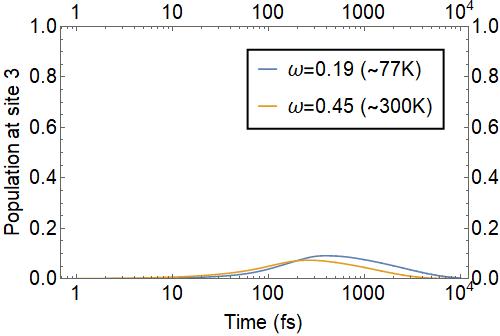}}
    %\label{}
\subfigure[]{    \includegraphics[width=.24\textwidth]{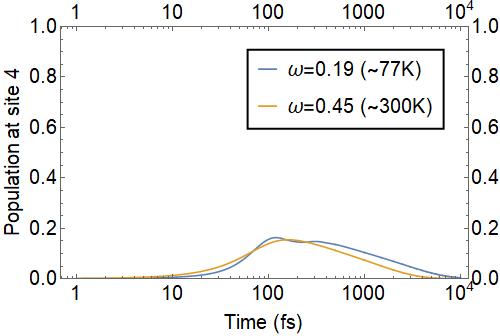}}
    %\label{}
\subfigure[]{    \includegraphics[width=.24\textwidth]{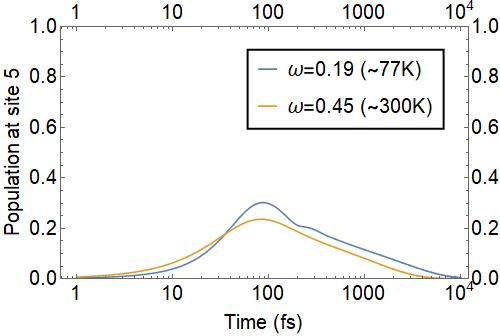}}
    %\label{}
\subfigure[]{    \includegraphics[width=.24\textwidth]{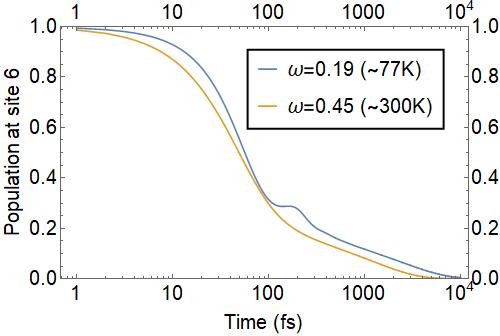}}
    %\label{}
\subfigure[]{    \includegraphics[width=.24\textwidth]{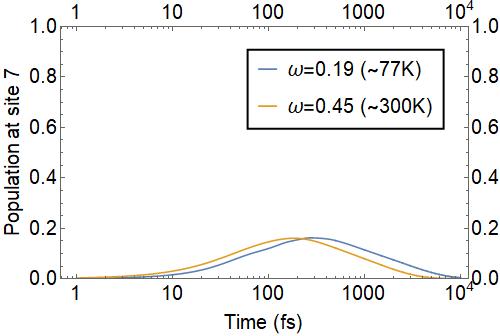}}
    %\label{}
\caption{Site population versus time for various sites at $\omega=0.19$ and $\omega=0.45$ for dephasing and incoherence model.}
\label{sitewisepopdeph+dec}
\end{figure*}
\subsection{Site population plots for QSW with pure dephasing}
Here, we have plotted (Fig.~\ref{sitewisepopdeph}) the evolution of exciton population at each site in the  QSW model with pure dephasing at $\omega=0.19$ and  $\omega=0.45$. We note from each of the plots Fig.~\ref{sitewisepopdeph}(a-g) that at $\omega=0.45$ the exciton population reaches the peak at a slower rate than at $\omega=0.19$. This leads to increase in the localization time for $\omega=0.45$ implying a speed up.
 \subsection{Comparison of QSW model with only incoherence and Hoyer's model at $77~K$}
 We have plotted in Fig.~\ref{sitewisepopdec} the evolution of the exciton population at different sites for QSW model with only incoherent scattering at $\omega=0.19$ ( temperature$77~K$) and the same for Hoyer, et. al.'s model at $77~K$. We also put the plots for the same at $300~K$ for comparison. As Hoyer, et. al.'s model did not have the plot for site $1$, we omit site $1$ from this analysis. To show the plot for the same at $77~K$, we have extracted data from Fig.~3(d) of Ref.~\cite{Hoyer_2010} using an online tool, WebPlotDigitizer~\cite{rohatgi2017webplotdigitizer}. We see that although the values are a bit different, the trend followed is similar in the plot for QSW with only incoherence and Hoyer's model at $77~K$. And since these plots do not coincide, this leads to different localization time. From Figs.~\ref{sitewisepopdec}(a-f) we see that the site population for each site in the only incoherence scheme at $\omega=0.19$ (temperature $77~K$) reaches a peak at a slightly earlier time as compared to Ref.~\cite{Hoyer_2010} at $77~K$. This agrees with the slightly lower localization time of $60.91 fs$ in case of QSW with only incoherence as compared to $70~fs$ in the case Hoyer's model. Also from Figs.~\ref{sitewisepopdec}(a-f) we see that for only incoherent scattering scheme at $\omega=0.45$ ( temperature $ 300~K$), the population at each site reaches a peak even earlier. This again agrees with the decreased localization time, which is $17.15 fs$ for QSW model with only incoherence at $300~K$. In conclusion, localization time is less at $300~K$ than at $77~K$ for QSW model with only incoherence. This explains the  major departure from Hoyer's model which has same localization time, that is $70~fs$ for both $77~K$ and $300~K$. A housekeeping note on the use of the WebPlotDigitizer~\cite{rohatgi2017webplotdigitizer} tool, that in Fig.~3(d) of Ref.~\cite{Hoyer_2010} the x-axis is in logarithmic scale. We extract the data points by manually clicking on the graph. WebPlotDigitzer interpolates and gives us the data points of the graph as a .csv file. 
% \begin{figure*}[!]
%\subfigure[]{    \includegraphics[width=.24\textwidth]{s1.pdf}}
    %\label{}
%\subfigure[]{    \includegraphics[width=.24\textwidth]{s2.pdf}}
   % \label{}
%\subfigure[]{    \includegraphics[width=.24\textwidth]{s3.pdf}}
    %\label{}
%\subfigure[]{    \includegraphics[width=.24\textwidth]{s4.pdf}}
    %\label{}
%\subfigure[]{    \includegraphics[width=.24\textwidth]{s5.pdf}}
    %\label{}
%\subfigure[]{    \includegraphics[width=.24\textwidth]{s6.pdf}}
    %\label{}
%\subfigure[]{    \includegraphics[width=.24\textwidth]{s7.pdf}}
    %\label{}
%\caption{Site population versus time for various sites at 77~K and 300~K for QSW model with both dephasing and incoherent scattering.}
%\end{figure*}
\subsection{Site population plots for QSW with incoherent scattering and dephasing}
Here, we have plotted (Fig.~\ref{sitewisepopdeph+dec}) the evolution of exciton population at each site in the  QSW model with both dephasing and incoherence at $\omega=0.19$ and  $\omega=0.45$. We note from each of the plots Fig.~\ref{sitewisepopdeph+dec}(a-g) that at $\omega=0.45$ the exciton population reaches the peak at a much faster rate than at $\omega=0.19$. This leads to decrease in the localization time for $\omega=0.45$ implying a slow down. Thus, rendering the QSW model with both dephasing and incoherence ineffective in explaining the possible quantum speed up expected in exciton transport in FMO complex.
\subsection{Correspondence between $\omega$ and temperature via total site coherence plots}
In order to get values of $\omega$ corresponding to particular temperatures, we juxtapose on top of each other the total site coherence plot from the QSW model with only incoherence with that from Hoyer's model~\cite{Hoyer_2010}. We have extracted the data from Fig.~3(c) of Ref.~\cite{Hoyer_2010} using the online tool, WebPlotDigitizer~\cite{rohatgi2017webplotdigitizer}. We then plot the extracted data and compare it with plot of total site coherence for various values of $\omega$. We find that for $\omega=0.19$ and $\omega=0.45$, the plots of total site coherence for QSW model with pure dephasing are most similar to the extracted plot at $77~K$ and $300~K$ from Fig.~3(c) of Ref.~\cite{Hoyer_2010}. But, there is a slight offset in the QSW plots from that of the plots of Ref.~\cite{Hoyer_2010}. The reason for such a shift is that although the initial total site coherence of both models are the same due to initial density matrices being the same, the master equations used in both models are different. In case of Hoyer, et. al., the master equation only has contribution from the Hamiltonian in the first term. On the other hand, in our case, since the Lindblad operators we use all depend on the Hamiltonian, the contribution from the Hamiltonian enters in all the terms of Eq.~(\ref{QWS_def}). Thus, the density matrices evolve differently in both models which is why we observe the offset.
%\begin{figure*}
%\centering
%\subfigure[]{ \includegraphics[width=.4\textwidth]{tsc_77.pdf}}
    %\label{}
%\subfigure[]{ \includegraphics[width=.4\textwidth]{tsc_300.pdf}}
    %\label{}
%\caption{Total site coherence versus time for (a) Hoyer's model\cite{Hoyer_2010} at temperature $T=77~K$ and QSW with pure dephasing for $\omega=0.19$ (b) Hoyer's model at temperature $T=300~K$ and  QSW with pure dephasing for  $\omega=0.45$}
%\label{total site coherence}
%\end{figure*}
\end{widetext}
\section{Data Availability Statement}
The authors confirm that the data supporting the findings of this study are available within the article and the Appendix.
\bibliographystyle{ieeetr}
\bibliography{ref-new}
\end{document}